\documentclass{article}
\usepackage{graphicx}
\usepackage{geometry}
\begin{document}

\title{SIR model on one dimensional small world networks} 
\author{M. Ali Saif$^1$\footnote{masali73@gmail.com}, M. A. Shukri$^2$\footnote{mshukri2006@gmail.com} and F. H. Al-makhedhi$^2$\footnote{fatymeheod@gmail.com}\\
$^1$ Department of Physics,
University of Amran,
Amran,Yemen\\
$^2$ Department of Physics, University of Sana'a, Sana'a, Yemen.}

\maketitle
\begin{abstract}
We study the absorbing phase transition for the model of epidemic spreading, Susceptible- Infected- Refractory (SIR), on one dimensional small world networks. This model has been found to be in the universality class of the dynamical percolation class, the mean field class corresponding to this model is $d=6$. The one dimensional case is special case of this class in which the percolation threshold goes to one (boundary value) in the thermodynamic limit. This behavior resembles slightly the behavior of one dimensional Ising and XY models where the critical thresholds for both models go to zero temperature (boundary value) in the thermodynamic limit. By analytical arguments and numerical simulations we demonstrate that, increasing the connectivity ($2k$) of this model on regular one dimensional lattice does not alter the criticality of this model. Whereas we find that, this model crosses from a one dimensional structure to mean field type for any finite value of the rewiring probability ($p$), in manner is similarly to what happened in the Ising and XY models on small world networks. In additional to that, we calculate the critical exponents and the full critical phase space of this model on small world network. We also introduce the crossover scaling function of this model from one dimensional behavior to mean field behavior. Furthermore we reveal the similarity between this model, and the Ising and XY models on the small world networks.


\end{abstract}
      

\section{Introduction}
Classification the systems which they have a phase transition from an active phase to an absorbing phase into universal classes has been attracted a lot of attention recently \cite{hen,hin,gez,mar}. The most celebrated class of this kind which is very robust respect to microscopic changes and has been studied extensively, was the directed percolation (DP) universality class. Systems display a continuous phase transition into a unique absorbing state with a positive one-component order parameter, with short-range interactions and without quenched disorder or additional symmetries were belong to this class \cite{hin,jans,gras}. DP class even has been found to be more general, some systems with long range interaction or certain models with nonunique or fluctuating passive states belong to this class \cite{hin,gez,she,alb,mun1}. Nevertheless, there were found models which undergo a phase transition to absorbing phase do not belong to this class. For example, the systems with two symmetric absorbing states belong to the voter models classes \cite{hen,hin,gez}. The systems in which the parity of the number of particles is conserved belong to parity-conserving classes\cite{hen,hin,gez}. The epidemic models that include immunization generically belong to the class so called dynamical percolation classes (DyP) \cite{hen,hin,gez,gra}. 

In this work we are going to study the phase transition from active phase to absorbing phase for the model of the spreading of an epidemic with immunization, i. e., Susceptible- Infected- Refractory (SIR), on one dimensional small world networks which some authors they call it a general epidemic process \cite{gra,car,jan0,car1,grass,janm,tom0}. This model can be mapped into the prey and predators model where the mapping can be accomplished by making the correspondence: susceptible and infected individuals with prey and predators individuals, respectively \cite{ant,ara}. Also this model can be interpreted as a limiting case of the forest fire model \cite{bak}, or also it is a special case of the Susceptible-Infected-Removed-Susceptible (SIRS) model \cite{sou}. SIR model assumes the individuals can be in a one state of the three states: Susceptible (S), Infected (I) and Recovered (R). The interaction in between the individuals for this model is as follows: a susceptible individual becomes infected with a given infection rate during contact with infected individuals, and an infected individual at given recovering rate becomes recovered. With this procedure, when an infected individual place in a community of susceptible individuals, initially and during a certain period of time the number of infected individuals increase. After that the infected individuals start decreasing due to immunization and the system eventually approaches a disease free state by means the individual become recovered or still susceptible. It is clear that, for small values of infection rate epidemic will spread locally within small numbers of susceptibilities individuals before it die out. However when the infection rate is large, there is a possibility for disease to spread over the enter system. Hence, the SIR model can exhibit a phase transition from the active phase (infection spreads) to adsorbing phase (non-spreading) at specific value of infection rate.

studying the phase transition of SIR model on the two dimensional regular network has been placed it in the dynamical percolation class \cite{gra}. DyP is a universality class of nonequilibrium phase transitions to absorbing states, which differs significantly from DP class. In the language of epidemic spreading, dynamical percolation class is most readily introduced as a generalization of directed percolation including the effect of immunization. Stauffer and Aharony \cite{sta} introduce geometrical interpretation of this model as ordinary percolation system (isotropic (undirected) percolation) describes the occurrence of infinitely large connected cluster of nearest neighbors sites in d-dimensions lattice, when each of those lattice sites is randomly and independently occupied with probability $P$. Generally, such percolating clusters can be generated by the dynamical percolation process. More specifically, whenever the epidemic spreading process is terminated, it leaves behind a region of immune sites. This cluster can be shown to be an ordinary isotropic percolation cluster. In this sense, this model will have infinitely many absorbing states. Using a field theoretical treatment Benzoni and Cardy \cite{ben} have shown that, The upper critical dimension of this class is $d_c=6$. However, the case of $d=1$ is special case in which the critical threshold is $P_c=1$ \cite{gez}. This behavior of DyP at a one dimension remind us the behavior of Ising model \cite{bar}, or even XY model \cite{jun} at the same dimension $d=1$ in which the two models do not have long-range order at any nonzero temperatures ($T_c\neq0$). Studying both models on a one dimensional small world network has been revealed that, as the long range interaction increases both models cross from a one dimension behavior to mean field type with the critical threshold changes as the strength of interaction changes \cite{bar,jun}. These results motive us to study the SIR model on small world network to reveal the behavior of this model under the effects of the long range interaction, and also to show if there are any kind of similarity between this model and Ising model and XY model on small world networks.

We should mention here also to that, SIR model has been studied with different modifications. Dammer and Hinrichsen \cite{dam} have studied the influence of a pathogen mutates during transmission of this model on a two dimensional lattice. They have shown that, mutation can drive the model form DyP class to DP class.  
Janssen et. al. \cite{jan} have considered this model with weak individuals, this modification leads to appearance a discontinuous transition in addition to the usual continuous percolation transition separated by the dynamic isotropic percolation universality class. Tome and Oliveira \cite{tom} explore the behavior of this model on a Cayley tree network. Dickison et. al. \cite{dic} studied SIR model on interconnected networks. The effects of dilution and mobility on the critical immunization rate of this model in two-dimensional lattices have been investigated by Silva and Fernandes \cite{sil}. Also SIR model with noise effect has been studied by Ji and Jiang \cite{ji}. On small world networks with shortcuts added Moore and Newman \cite{moo} have found that, the density of infected individuals are a function of the density of shortcuts.    
 
\section{Model and Methods}
The SIR model of epidemic spreading on the networks, can be described as follows \cite{gra}: consider a population of $N$ individuals live at the sites of a one dimensional lattice. The individual can be in one state of three states, susceptible $(S)$, infected $(I)$ and refractory $(R)$. Individuals in state $I$ on the network can infect any one of their neighbors which are in state $S$ with infection probability $\lambda$ at each time step. The individuals in state $I$ pass to the $R$ state with recovering probability is one (we follow Grassberger's second model in Ref. \cite{gra}). During the $R$ phase, the individuals are immune and do not infect. It is evident that, this model will evolve ultimately to the state where there are no infected individuals in the lattice, that is for any initial state of $I$ individuals and for any values of $\lambda$. That means the final state is mixture of $S$ and $R$ individuals. Hence this state is the absorbing state of the model. The order parameter considered here is the density of individuals $\left\langle \rho\right\rangle$ in the phase $R$ averaged over different network realizations.

We generate the small world network from a regular one dimensional network using Watts and Strogatz (WS) \cite{wat} method. In WS small world network, we start from a regular network with $N$ nodes in which every node in the network is connected to its $2k$ nearest neighbors. Consequentially, each link on the network removed with rewiring probability $p$ and reconnected to a randomly chosen site. Therefore, the case when $p=0$ corresponds a regular network, however the case when $p=1$ represents a random graph.

At this point we will try here to extract some useful relations related to this model. And for simplicity, we consider the regular one dimensional long chain with $N$ individuals Fig. 1. In this chain each individual connects to its nearest neighbor as shown in the figure. If we suppose the infection starts at the left most individual on the chain. This individual can transmit the infection to its neighbor with probability $\lambda$ and the new infected individual can also transmit the diseases to its neighbor with probability $\lambda$ and so on. Thus, the probability $P_n$ of surviving such that process for $n-1$ trials and ultimately terminates is $P_n=\lambda^{n-1} (1-\lambda)$ \cite{rei}. This probability is properly normalized $\sum^{n=1}_{\infty} P_n=1$. Hence, depending on the value of $\lambda$ the mean number of infected individuals before the process is terminated is \cite{rei}:  
\begin{eqnarray}
\bar{n}=\frac{1}{1-\lambda}   
\end{eqnarray}  

With this process the percolation threshold attains when all sites of the chain become infected $I$. For every value of $\lambda<1$, there will be some individuals in the chain in the state $S$. Thus the percolation threshold is $\lambda_c=1$ \cite{sta}. The density of infected individuals during this process is $\left\langle \rho\right\rangle=\frac{\bar{n}}{N}$

\begin{figure}[htb]
 \includegraphics[width=60mm,height=30mm]{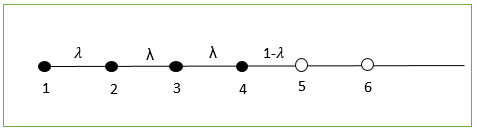}
\caption{Example of a one dimensional chain in which the infection starts at the left most site and the infection propagates with probability $\lambda$ through the sites of the chain.}
 \end{figure} 

The general case is the case when each individual in the lattice is connected to more than one of its neighbors i. e., the case when $k>1$. In this case, that individual in the network can be infected by more than one of its neighbors, where there are $2k$ neighbors for each individual. Therefore, the probability of infection will be proportional to the mean connectivity of the network $\left\langle k\right\rangle$. Therefore, we expect that probability will be given as follows $\pi= 1-(1-\lambda)^\gamma$ \cite{car1,kup,yan} and the value of $\gamma$ to be $\gamma\approx \left\langle k\right\rangle$. In this case the surviving probability is $P_n=\pi^{n-1} (1-\pi)$ and the mean number of infected individuals is $\bar{n}=\frac{1}{1-\pi}$ or:
\begin{eqnarray}
\bar{n_k}=\frac{1}{(1-\lambda)^\gamma}   
\end{eqnarray}

Density of infected individuals will be given again as follows
\begin{eqnarray}
\left\langle \rho\right\rangle=\frac{\bar{n_k}}{N}
\end{eqnarray}  

It is clear also here that, the value of the critical threshold in thermodynamic limit is $\lambda_c=1$.

For the purpose of calculating the order parameter $\left\langle \rho\right\rangle$ numerically, we perform Monte Carlo simulations for the SIR model on regular one dimensional for different values of $N$ and at various values of the systems's parameter $\lambda$. The network we consider it here is a periodic and the system updates are synchronously. We fixed the value of $k$ to be $k=3$ for all calculations we carry out in this work. Fig. 2, shows the simulation results for the density of infected individuals $\left\langle \rho\right\rangle$ for the values of $N=2\times 10^3, 5\times 10^3, 10\times 10^3$ and $20\times 10^3$ (from left to right) at the various values of the system's parameter $\lambda$. Each point in the figure has been averaged over $10000$ independent runs for $N=2\times 10^3, 5\times 10^3, 10\times 10^3$ and $5000$ independent runs for $N=20\times 10^3$. This figure shows clearly that, the finite size effects affect strongly on this model on one regular network and as it is clear as the values of $N$ increase the curves shift toward right. This behavior suggests that, the critical threshold will move toward one, i. e. $\lambda_c=1$, as $N\rightarrow \infty$. In Fig. 2 the corresponding solid curves to each value of $N$ are due to the scaling function Eq. 3. For the best fit we find the value of $\gamma$ in Eq. 2 to be $\gamma=6.3(2)$. This value diverts slightly from the expected value of $\gamma=\left\langle k\right\rangle=2k=6$ for this case.
\begin{figure}[htb]
 \includegraphics[width=70mm,height=60mm]{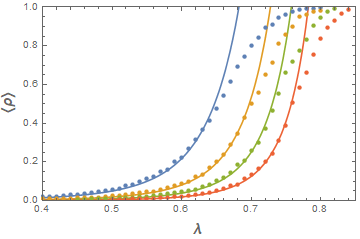}
\caption{Density of infected individuals $\left\langle \rho\right\rangle$ as function of $\lambda$ for $N=2\times 10^3, 5\times 10^3, 10\times 10^3$ and $20\times 10^3$ from left to right for one dimensional regular lattice ($p=0$). The corresponding solid curves to each value of $N$ are obtained from scaling Eq. 3 with $\gamma=6.3$.}
 \end{figure} 

For comparison with the small world networks, we plot in Fig. 3 the calculated values for the order parameter $\left\langle \rho\right\rangle$ as function of $\lambda$ for the values of $N=2\times 10^3$ and $10\times 10^3$ when $k=3$ and the rewiring probability is $p=1$. As the figure shows, the two curves of $N=2\times 10^3$ and $N=10\times 10^3$ collapse on each other unlike the previous case when $p=0$ (where the curves go apart). This behavior indicates to existence the critical thresholds for SIR model under the effects of long range connection at the values of $\lambda$ are less than one. As we show in the next section, for the best estimates and when $p=1$ the critical point is $\lambda_c=0.176(2)$ when $N=10^4$, however it is $\lambda_c=0.175(2)$ when $N=10^5$. Thus, the finite size effects do not play here an important role in contrast with the regular lattice.  
\begin{figure}[htb]
 \includegraphics[width=70mm,height=60mm]{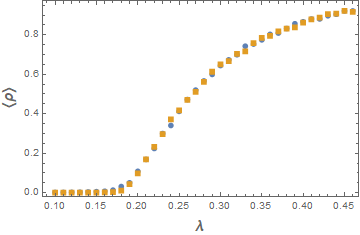}
\caption{Density of infected individuals $\left\langle \rho\right\rangle$ as function of $\lambda$ for $N=2\times 10^3$ and $10\times 10^3$ when $p=1$.}
 \end{figure}

\section{Time-dependent simulation}
The most accurate technique to study the critical properties of any system with absorbing states comes from the so-called dynamical Monte Carlo method \cite{gra1,jens,dicm}. Using this technique we determine the critical points and extract the critical exponents associate to the system under consideration and identify its universal class. In this method, the time evolution of the system starts initially from the one active individual at the origin of a nonactive environment. Considering many configurations of this system start from the same initial state, we can calculate for example the following quantities which exhibit a power law behavior when the system is critical. Those quantities are the survival probability $P(t)$ and the number of active sites $N(t)$ averaged over all independent runs. At the critical point these quantities decay asymptotically according to a power law
\begin{eqnarray}
P(t)\sim t^{-\delta} 
\end{eqnarray} 
and
\begin{eqnarray}
N(t)\sim t^\theta
\end{eqnarray} 
Using the previous equations we can determine the values of the critical exponents $\delta$ and $\theta$. However, the efficient and systematic way to estimate those critical exponents as well as the critical points from Eqs. 4, 5 is to analyze the local slope method by introducing the effective exponent defined as follows \cite{hin,gra3}:
\begin{eqnarray}
-\delta(t)=\frac{\ln[P(t)]-\ln[P(t/m)]}{\ln[m]}
\end{eqnarray}
where $m$ is a constant larger than one. Similar definition we can find for the effective exponent of $\theta(t)$. 

In additional to that, we also estimate the critical exponent corresponding to the time correlation length $\nu_{\parallel}$ using the time evolution of the derivative $D$ of the mean value of the survival probability $\left\langle P\right\rangle$ which is given by the relation \cite{mar,gra2}:
\begin{eqnarray}
D=\frac{d \log\left\langle P\right\rangle}{d \log \lambda}\sim t^{1/\nu_{\parallel}}
\end{eqnarray}

In order to evaluate the critical points of this model and also the values of the critical exponents in Eqs. 4-6, we perform Monte Carlo Simulations of SIR model at various values of the rewiring probability $p$. We have fixed the value of $k$ to be $k=3$ for all cases we have studied. We believe same conclusion should hold for any values of $k>1$. Our Monte Carlo simulations for regular one dimensional lattice ($p=0$) of this model, do not show any kind of power law at any value of the system's parameter $\lambda$ and for any finite values of $N$. This behavior confirms what we have found in the previous section in which the critical point of this model for the case when $p=0$ is $\lambda_c=1$, unlike the case when $p=1$ where the system shows a well defined power law behavior at a specific value of the model's parameter $\lambda$. 

We show in Figs. 4, 5 and 6 the results of Monte Carlo simulations for the quantities, the survival probability $P(t)$, the number of active sites $N(t)$ and the derivative $D$ for some selected values of $p$. From those figures and for example at the value of $p=0.5$, the critical point is $\lambda_c=0.190(2)$. The estimated values of the critical exponents are $\delta=1.00(2)$, $\theta=0.00(1)$ and $\nu_{\parallel}=1.04(4)$. Those values coincide very well with the values of critical exponents for the mean field of DyP universality class \cite{sta,mun,hen}. This means that, the phase transition of SIR model on WS small world networks is of kind DyP mean field universality class. Particularly, for any value of $p$ the critical points $\lambda_c$ of this model have been determined to be the value of $\lambda$ in which both of $P(t)$ and $N(t)$ display a power law behavior as implied by scaling functions Eqs. 4, 5. 

In the table 1 and using the same process described above, we summarize the estimated values of the critical exponents and the critical points derived from our Monte Carlo simulations for various values of rewiring probability $p$. It is clear that, for all values of $p$ which we have studied, the critical exponents those we find them are in good agreement with the values of the critical exponents of mean field of DyP class \cite{sta,mun,hen}. Those results confirm that, the SIR model crosses from one dimensional system to the mean field like system under the effect of the long range interaction, in similar manner to what happened in the one dimensional WS small world Ising model \cite{bar}, and the XY model on one dimensional WS small world networks \cite{jun}. 

\begin{figure}[htb]
 \includegraphics[width=70mm,height=60mm]{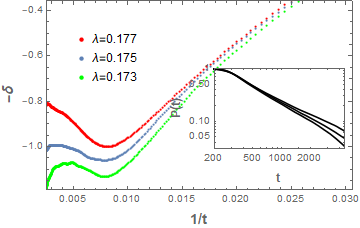}
 \includegraphics[width=70mm,height=60mm]{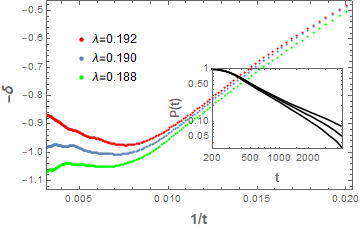}
\caption{Time dependent behavior of the effective exponent for the Survival probability exponent $\delta$ as function of $1/t$ for $p=1.0$ (left) and $p=0.5$ (right) for lattice size $N=10^5$ and $k=3$. Inset shows the survival probability $P(t)$ as function of time. The data are averaged over $10^4$ realization.}
 \end{figure} 

\begin{figure}[htb]
 \includegraphics[width=70mm,height=60mm]{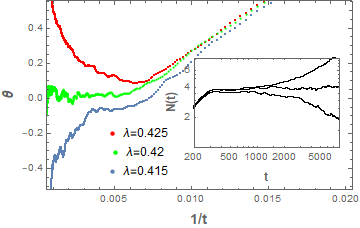}
 \includegraphics[width=70mm,height=60mm]{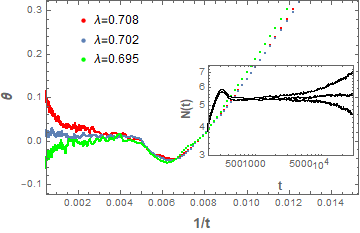}
\caption{Time dependent behavior of the effective exponent for the number of active sites exponent $\theta$ as function of $1/t$ for $p=0.01$ (left) and $p=0.0001$ (right) for lattice size $N=10^5$ and $k=3$. Inset shows the number of active sites $N(t)$ as function of time. The data are averaged over $10^4$ realization.}
 \end{figure}   

\begin{figure}[htb]
 \includegraphics[width=70mm,height=60mm]{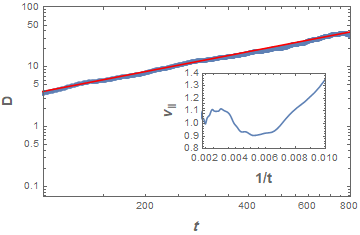}
 \includegraphics[width=70mm,height=60mm]{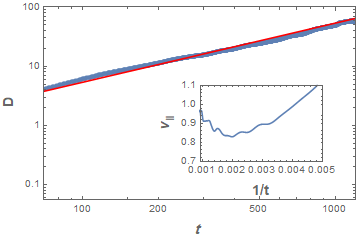}
\caption{Time dependent behavior of the derivative $D$ of the mean value of the survival probability $\left\langle P\right\rangle$ as function of $t$ for $p=0.5$ (left) and $p=0.08$ (right) for lattice size $N=10^5$ and $k=3$. Inset shows the critical exponent of $\nu_{\parallel}$ as function of $1/t$. The data are averaged over $10^4$ realization.}
 \end{figure}   

\begin{tabular}{ccccc}
\multicolumn{5}{l}{Table 1: Critical points and Critical exponents.}\\
\hline
p & $\lambda_c$ & $\delta$ & $\theta$ & $\nu_{\parallel}$\\
\hline
1.0& 0.175(2) & 1.01(2) & 0.01(2)& 1.11(5)\\
0.5& 0.190(2) & 1.00(2) & 0.00(1)& 1.04(4)\\
0.3& 0.208(2) & 0.99(3) & 0.00(1)& 1.10(4)\\
0.2& 0.227(2) & 0.99(3) & 0.00(2)& 1.10(4)\\
0.1& 0.264(2) & 0.98(3) & 0.00(3)& 1.10(4)\\
0.05& 0.307(3) & 0.98(4) & 0.01(3)& 1.10(4)\\
0.03& 0.341(3) & 0.96(5) & 0.01(3)& 1.08(4)\\
0.02& 0.370(3) & 0.97(5) & 0.02(4)& 1.07(4)\\
0.01& 0.420(4) & 0.97(5) & 0.03(4)& 1.08(4)\\
0.005& 0.467(4) & 0.97(6) & 0.03(4)& 1.06(4)\\
0.002& 0.530(4) & 0.95(6) & 0.02(4)& 1.06(4)\\
0.001& 0.575(4) & 0.95(6) & 0.03(4)& 1.07(5)\\
0.0005& 0.614(5) & 0.95(6) & 0.02(3)& 1.04(6)\\
0.0001& 0.703(5) & 0.95(6) & 0.02(3)& 1.03(7)\\
\hline
\end{tabular}\\

Fig. 7 shows the phase diagram $(p, \lambda_c)$ of the SIR model on small world networks with $k=3$. We find that, the critical threshold $\lambda_c$ of this model as function of the rewiring probability $p$ is very well described by the logarithmic scaling form: $\lambda_c(p)\sim 0.12-0.065\ln(p)$ for any values of $p\leq 0.2$ as that shown in the inset of Fig. 7. The scaling form we find here resembles the scaling form that has been found in XY model on one dimensional small world networks \cite{jun}. However to what limit of the values of $p$ our form is valid looks like to the scaling form that has been found in Ising model on one dimensional small world networks \cite{bar}, unlike the XY model where the scaling form was valid for all values of $p$ except for $p=1$. Extrapolation of our scaling form predicts a so small value
for the critical rewiring probability $p_c\approx 10^{-6}$. This supports the assumption of mean field type of this model for any nonzero of rewiring probability ($p>0$).

\begin{figure}
		\includegraphics{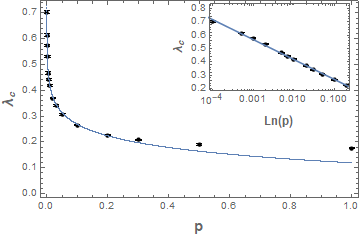}
	\caption{Schematic phase diagram of SIR model on small world networks. The values of the critical points $\lambda_c$ are plotted as function $p$ when $k=3$. The solid line is obtained from the scaling form: $\lambda_c(p)\sim 0.12-0.065\ln(p)$. Inset: The critical points are well described by the form $\lambda_c(p)\sim 0.12-0.065\ln(p)$ for any values of $p\leq 0.2$.}
\end{figure}

For more details, we re-ask the same question given in the Ref. \cite{jun}, what is the minimum value of $p$ at which the SIR model crosses from a one dimensional behavior to mean field one?. Answering of this question needs from us to understand the dependence of $p$ on the transition probability $\lambda_c$. In fact we have a two length scale characterize this system on small world networks. The first one comes from the small world networks, where the typical distance between ends of shortcuts is given by $\zeta=(kp)^{-1}$ \cite{new}. The other one we can get it from the correlation length of the pure SIR model on one dimensional regular lattice which is given by $\xi\sim \frac{-1}{\ln(\pi)}$ \cite{sta}, where $\pi= 1-(1-\lambda)^\gamma$. Therefore, when the correlation length $\xi$ is smaller than the typical distance between ends of shortcuts $\zeta$, the system behaves fundamentally as a regular network. On the other hand when the correlation length $\xi$ grows beyond $\zeta$, the long range interaction due to the shortcuts come into play, inducing the mean field behavior. Consequently, the crossover from the one dimensional behaviour to mean field type will happen when $\xi \approx \zeta$ \cite{bar,jun,her}. Thus for fixed values of $\lambda$ the SIR model crosses from one dimensional behavior for $p<p_{co}$, where $p_{co}$ is,  
\begin{eqnarray}
p_{co}(\lambda)\propto \frac{-\ln(1-(1-\lambda)^\gamma)}{k} 
\end{eqnarray} 
  
to mean field for larger $p$.
On the other way, for fixed values of $p$ the crossover from one dimensional to mean field will happen when $\lambda>\lambda_{co}$ where $\lambda_{co}$ is
\begin{eqnarray}
\lambda_{co}(p)\propto 1-(1-\exp(-kp))^{1/\gamma}
\end{eqnarray}

In Fig. 8 we replot the simulation results in Fig. 7 for critical points $\lambda_c$ as function of rewiring probability $p$ when $k=3$ along with the asymptotic scaling form Eq. 9 for $k=3$ (solid line) and $k=4$ (dashed line) for comparison. We consider the value of $\gamma$ to be equal to $2k$. The figure shows clearly that, the critical points are higher than the crossover line (solid line for $k=3$), which means the critical points are located at mean field region. The distance from the crossover line to the critical points is large for high values of $p$ which confirms the existing of mean field behavior at least for sufficiently large values of $p$, however that distance starts decreasing as the values of $p$ is very small. In our Monte Carlo simulation we have demonstrated the existing of the mean field behavior of this model up to very small values of $p=10^{-4}$, which is likely to suggest $p_c=0$ in the thermodynamic limit. In additional to that, Eq. 9 suggests the existence of crossover threshold for any value of $p>0$. However, as it is clear from Fig. 8 at very small values of $p$ the critical points $\lambda_c$ are very close from the crossover line, in which becomes the determination of critical points and critical exponents accurately needs to employ larger networks.  
\begin{figure}
		\includegraphics{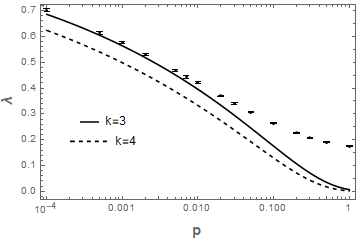}
	\caption{The critical points as function of $p$ as in Fig. 7. By scaling form Eq. 9 the solid line represents the crossover from one-dimensional to mean field behavior when $k=3$. Dashed line is crossover line when $k=4$.}
\end{figure}

Finally with same fashion in Refs. \cite{bar,jun,her}, we introduce an approximate calculation for the critical points of this model when $p<<1$. In fact, for very small values of $p$ the typical distance of small world networks is $\zeta\sim p^{-1}$ \cite{jun,her}, however we can approximate the correlation length of our model to be $\xi\sim -(\ln \lambda)^{-1}$ \cite{sta}. Therefore when $p<<1$, we expect the critical behavior of this system to behave as $\lambda_c\sim \exp(-p)$. This equation predicts that, the critical point $\lambda_c$ will go to one (one dimensional behavior) as the value of $p\rightarrow 0$. For comparison with Ising model on small world networks and for very small values of $p$ the critical threshold was given $T_c\sim -(\ln p)^{-1}$ \cite{bar}. However the critical threshold in the XY model was given by $T_c\sim p$ \cite{jun}. The reason behind this diversity in the critical thresholds for the three models backs to the different in the correlation length for each model. Whereas the correlation length for SIR model is given by $\xi\sim -(\ln \lambda)^{-1}$ \cite{sta}, the correlation length of Ising model is exponentially depend on the temperature as follows $\xi\sim \exp(a/T)$ where $a$ is a constant \cite{bar}. However, the correlation length for XY model is given by $\xi\sim T^{-1}$ \cite{nag}.

\section{Conclusion}
We have studied the SIR model for epidemic spreading on one dimensional small world networks. In the one dimensional regular chain this model does not show a phase transition where the critical threshold is $\lambda_c=1$ in the similar behavior to the Ising and XY models on the same dimension. In small world networks we have demonstrated that, this model crosses from one dimensional behavior to mean field type for any small values of rewiring probability $p$. Explicitly, we have checked the mean field type of this model with Monte Carlo simulations up to small values of $p$. we have also inferred the crossover scaling function from one dimensional to mean field of this model and extracted the dependent of the critical threshold $\lambda_c$ on the rewiring probability $p$. Our results suggest the mean field type of the model for any values of $p\neq0$ in the thermodynamic limit. This behavior seems to be related to the small world phase transition, where the characteristic path length ${\it l}$ shows a change of behavior for any nonzero values of $p$ in the thermodynamic limit \cite{new1,bart,dem}.

 \end{document}